\documentclass[12pt]{article}
\usepackage{graphicx}
\usepackage{epsfig}
\usepackage{setspace}
\usepackage{amsmath}

\begin{document}

\title{Phase transitions of barotropic flow coupled to a massive rotating sphere -
derivation of a fixed point equation by the Bragg method}
\author{Chjan C. Lim and Rajinder Singh Mavi \\
%EndAName
Mathematical Sciences, RPI, Troy, NY 12180, USA\\
email: limc@rpi.edu}
\date{\today }
\maketitle

\begin{center}
\textit{{Abstract} }
\end{center}

The kinetic energy of barotropic flow coupled to an infnitely massive
rotating sphere by an unresolved complex torque mechanism is approximated by
a discrete spin-lattice model of fluid vorticity on a rotating sphere,
analogous to a one-step renormalized Ising model on a sphere with global
interactions. The constrained energy functional is a function of spin-spin
coupling and spin coupling with the rotation of the sphere. A mean field
approximation similar to the Curie-Weiss theory, modeled after that used by
Bragg and Williams to treat a two dimensional Ising model of ferromagnetism,
is used to find the barotropic vorticity states at thermal equilibrium for
given temperature and rotational frequency of the sphere. A fixed point
equation for the most probable barotropic flow state is one of the main
results.

This provides a crude model of super and sub-rotating planetary atmospheres
in which the barotropic flow can be considered to be the vertically averaged
rotating stratified atmosphere and where a key order parameter is the
changeable amount of angular momentum in the barotropic fluid. Using the
crudest two domains partition of the resulting fixed point equation, we find
that in positive temperatures associated with low energy flows, for fixed
planetary spin larger than $\Omega _c>0$ there is a continuous transition
from a disordered state in higher temperatures to a counter-rotating
solid-body flow state in lower positive temperatures. The most probable
state is a weakly counter-rotating mixed state for all positive temperatures
when planetary spin is smaller than $\Omega _c.$

For sufficiently large spins $\Omega >2\Omega _c$, there is a single smooth change
from slightly pro-rotating mixed states to a strongly pro-rotating ordered
state as the negative value of $T$ increases (or decreases in absolute
value). But for smaller spins $\Omega <2\Omega _c$ there is a
transition from a predominantly mixed state (for $T\ll -1)$ to a
pro-rotating state at $\beta ^{-}(\Omega )<0$ plus a second
$\beta _c^\Omega,$ for which the fixed point
equation has three fixed points when $\beta <\beta _c^\Omega $ instead of
just the pro-rotating one when $\beta _c^\Omega <\beta .$ An argument based
on comparing free energy show that the pro-rotating state is preferred when
there are three fixed points because it has the highest free energy - at
negative temperatures the thermodynamically stable state is the one with the
maximum free energy.

In the non-rotating case ($\Omega =0)$ the most probable state changes from
a mixed state for all positive and large absolute-valued negative
temperatures to an ordered state of solid-body flow at small absolute-valued
negative temperatures through a standard symmetry-breaking second order
phase transition. The predictions of this model for the non-rotating problem
and the rotating problem agrees with the
predictions of the simple mean field model and the spherical model.

This model differs from previous mean field theories for quasi-2D turbulence
in not fixing angular momentum nor relative enstrophy - a property which
increases its applicability to coupled fluid-sphere systems and by extension
to 2d turbulent flows in complex domains such as no-slip square boundaries
where only the total circulation are fixed - as opposed to classical
statistical equilibrium models such as the vortex gas model and
Miller-Robert theories that fix all the vorticity moments. Furthermore, this
Bragg mean field theory is well-defined for all positive and negative
temperatures unlike the classical energy-enstrophy theories.

\newpage

\section{Introduction}

Previous statistical equilibrium studies of inviscid flows on a rotating
sphere with trivial topography focussed on the classical energy-enstrophy
formulation \cite{Kraichnan} of PDE models such as in \cite{FS1} did not
find any phase transitions \cite{Stanley} to super-rotating solid-body
flows. The main reason is that these Hamiltonian PDE models which conserve
the angular momentum of the fluid, are not suited to the study of a
phenomena like the super-rotation of the Venusian middle atmosphere that
depends on transfer of angular momentum between the atmosphere and the solid
planet.

We therefore consider instead a single layer of barotropic fluid of fixed
height coupled by a complex torque mechanism to a massive rotating sphere as
a simplified model of planetary atmosphere. The approach in this paper is to
consider different possible thermodynamic properties of the steady states of
coupled planetary atmospheres as mean field statistical equilibria - through
a natural relation of fluid vorticity systems to electromagnetic models -
with the aim of showing explicitly that sub-rotating and super-rotating flow
states can arise spontaneously from random initial states.

A simple mean field method on a related model \cite{mean}, (where spins are
allowed to take on a continuous range of values), has successfully found
phase transitions for barotropic flows coupled to a rotating sphere. There
as here, angular momentum of the fluid is allowed to change; and the
expected values of total circulation and relative enstrophy are fixed in
this simple mean field theory. This leads to the exciting result that
sub-rotational and super-rotational macrostates are preferred in certain
thermodynamic regimes. The Bragg mean field theory presented in this paper
corroborates some of these previous results but has the important new
property that relative enstrophy is constrained only by an inequality while
total circulation is held fixed at zero. Further discussion of these
important points will follow in the next section.

Our results are also in agreement with Monte Carlo simulations on the
logarithmic spherical model in the non-rotating case \cite{limnebus} and the
rotating sphere \cite{Dinglim}, \cite{Dinglim2}. The phrase spherical model
refers to the fact that under a discrete approximation, the microcanonical
constraint on the relative enstrophy becomes a spherical constraint first
introduced by Kac \cite{Stanley}. We shift gears here from the spherical
model used by Lim - for which closed form solutions of the partition
function required considerable analytical effort \cite{Limsphere06}, \cite
{iutam06} - to a simplified discrete spin model which enables us to use the
Bragg method to approximate the free energy, which in turn allows us to find
an interesting analytical solution to the coarse grained stream function for
the resulting mean field theory. While the overall qualitative agreement
between this Bragg model, Lim's exact solution of the spherical model and
the simple mean field model mentioned above is good, there are subtle
differences in the detailed predictions of these models. They are however
all within the scope of applications to super and sub-rotations in planetary
atmospheres such as on Venus, chiefly because current observations are not
refined enough to distinguish between them. It will be necessary to perform
detailed direct numerical simulations of coupled geophysical flows in order
to gather enough data for a future comparative study which can choose
between these three models. Nonetheless, we will discuss in the next section
how purely theoretical considerations of the complex boundary conditions
that exist in the coupled barotropic fluid -sphere system and in a class of
quasi-2d turbulent flows in complex domains allow us to choose the current
set of spin-lattice / Eulerian vorticity models with minimal constraints
over the vortex gas / Lagrangian models \cite{onsager}, \cite{Lundgren} and
Miller-Robert theories \cite{Miller}, \cite{Robert2}.

Our discrete vorticity model is a set $\mathbf{{X}=}$ $\left\{ \vec{x_i}%
\right\} $ of $2N$ random sites on the sphere with uniform distribution,
each with a spin $s_i\in \left\{ +1,-1\right\} $, and interaction energy
with every other site as a function of distance. Each spin also interacts
with the planetary spin which is analogous to an external magnetic field in
the Ising model - the sum of this interaction is proportional to the varying
net angular momentum of the fluid relative to a frame that is rotating at
the angular velocity of the solid sphere. Contribution to the kinetic energy
from planetary spin varies zonally, however, and this makes the external
field inhomogeneous and difficult to deal with analytically. Bragg and
Williams \cite{huang} used a one step renormalization to investigate
properties of order-disorder in the Ising Model of a ferromagnet. As the
discretized model for barotropic flows coupled to a rotating sphere is
similar to the Ising Model of a ferromagnet in an inhomogeneous field, we
use the Bragg-Williams renormalization technique to infer the order-disorder
properties of the fluid.

It is a well established fact \cite{onsager} that systems of constrained
vortices have both positive and negative temperatures, and the present model
is no exception. In this setup, using the simplest two domains partition of
the surface of the sphere, we find a positive temperature continuous phase
transition to the sub-rotating ordered state for decreasing temperatures if
the planetary spin is large enough. There is no evidence of a phase
transition for a non-rotating sphere in positive temperature. We also find 
transitions to a super-rotating ordered state when the negative temperature has
small absolute values - that is for very high energy.

\section{Statistical mechanics of macroscopic flows}

Here we summarize some of the pertinent points concerning the application of
a statistical equilibrium approach to macroscopic flows. The largely 2-d
flows we are concerned with in this paper are in actuality non-equilibrium
phenomena even in the case of nearly inviscid flows - an assumption that is
mainly correct for the interior of geophysical flows. The main reason that
equilibrium statistical mechanics is applicable here - the existence of two
separate time scales in the microdynamics of 2D vorticity - is best
formulated as the well-known physical principle of selective decay which
states that the slow time scale given by the overall decay rates of
enstrophy and kinetic energy in damped unforced flows is sufficiently
distinct from the fast time scale corresponding to the inverse cascade
relaxation of kinetic energy from small to large spatial scales so that
several relaxation periods fits in a unit of slow time. That means that the
total kinetic energy (and enstrophy) may be considered fixed in the time it
takes for the eddies to reach statistical equilibrium. Furthermore, the
principle of selective decay states that the asymptotic properties of the
damped 2d flow is characterized by a minimal enstrophy to energy ratio which
is dependent on the geometry of the flow domain. One of the key properties
of enstrophy known as the square-norm of the vorticity field then implies
that these minimum enstrophy states are associated with large scale ordered
structures such as domain scale vortices. In the case of the specific
problem in this paper, these coherent structures are super- and sub-rotating
solid-body flow states.

Aside from the above general points on the applicability of equilibrium
statistical mechanics to 2D macroscopic flows, the specific properties of a
given flow problem \cite{heijst} will decide which one of the many
statistical equilibrium models \cite{onsager}, \cite{Lundgren}, \cite
{Kraichnan}, \cite{Miller}, \cite{Robert2} is suitable. The coupled nature
of the flow problem in this paper where a complex torque mechanism transfers
angular momentum and energy between the fluid and rotating solid sphere,
clearly eliminates all Lagrangian or vortex gas models \cite{onsager}, \cite
{Lundgren} on the surface $S^2$ because they conserve angular momentum - a
result that follows directly from Noether's theorem. Moreover, since none of
the vorticity moments are conserved in the coupled flows except for total
circulation which is fixed at zero by Stokes theorem, independently (or in
spite) of the coupling between fluid and solid sphere, the Miller-Robert
class of models \cite{Miller}, \cite{Robert2} cannot be used here because
they are designed to fix all the vorticity moments. It is therefore not
surprising that a significant discrepancy was found between statistical
equilibrium predictions based on a vortex gas theory and experimental (and
direct numerical simulations) of decaying quasi-2D flows in a rectangular
box with no-slip boundary conditions \cite{heijst}.

What remains are the classical Kraichnan models - also known as absolute
equilibrium models - and variants such as those used in \cite{FS1}. The
problem with these are two-fold - (i) they model a 2d fluid flow that is
uncoupled to any boundary or has periodic boundaries and hence angular
momentum is held fixed - clearly unsuitable for our aims here, and (ii) they
are equivalent to the ubiquitous Gaussian model and can be easily shown to
not have well-defined partition functions at low temperatures. One solution
taken up in \cite{Limsphere06}, \cite{iutam06} is to impose instead of a
canonical enstrophy constraint, a microcanonical one which leads to a
version of Kac's spherical model. Combined with zero total circulation and
allowing angular momentum to fluctuate - that is coupling the barotropic
fluid to an infinitely massive rotating solid sphere- this approach yielded
exact partition functions and closed form expressions for phase transitions
to self-organized (or condensed) super- and sub-rotating flows. As expected,
these analytic results agree perfectly with Monte-Carlo simulations of the
spherical model in \cite{limnebus}, \cite{Dinglim}, \cite{Dinglim2}. An
argument similar in style to the above discussion of the Principle of
Selective Decay was needed to justify the choice of a microcanonical (and
hence spherical) constraint on relative enstrophy in this class of coupled
flows while the zero total circulation constraint needs none since it is
implied by topological arguments - Stokes theorem on the sphere. The
spherical model unlike the Bragg model in this paper is not based on a mean
field assumption. This can be a significant advantage only in modelling more
complex geophysical flows where large deviations techniques cannot be used
to establish the asymptotic exactness of the mean field.

One of the aims of this paper is therefore to relax the enstrophy constraint
- that is, we impose neither a canonical enstrophy constraint like in \cite
{Kraichnan} nor a microcanonical enstrophy constraint like in \cite
{Limsphere06}, \cite{iutam06}, \cite{limnebus}, \cite{Dinglim}, \cite
{Dinglim2} - and still derive a physically sound mean field theory for a
comparatively simple geophysical problem. We will show that the relative
enstrophy in the Bragg mean field theory is constrained instead by an upper
bound.

Another aim in this paper is to show that the Bragg method can be extended
to the situation as in this coupled flow where the kinetic energy is just a
functional and not a Hamiltonian of the barotropic flow. This is a subtle
but relevant point because the original Bragg method formulated for the
Ising model of ferromagnetism worked with a Hamiltonian. The first author
reported in \cite{iutam06} and \cite{Dinglim}, \cite{Dinglim2} that the
kinetic energy of the fluid component of the coupled fluid - rotating sphere
system cannot be a Hamiltonian for the evolution of the vorticity field
because otherwise its $SO(2)$ symmetry - a property that is easily shown -
implies the conservation of angular momentum of the fluid component.

\section{Coupled Barotropic Fluid - Rotating Sphere model}

We are interested in the physical possibility of planetary atmosphere at
thermal equilibrium with the rotating solid planet - the massive sphere is
treated as an infinite reservoir for both kinetic energy and angular
momentum of the barotropic (vertically averaged) component of the fluid.
Thus we consider the atmosphere as a single layer of incompressible inviscid
fluid interacting through an unresolved complex torque mechanism with the
solid sphere. This is equivalent to formulating a Gibbs canonical ensemble
in the energy of the fluid with an implicit canonical constraint on the
angular momentum of the fluid. More realistic models include effects of
forcing and damping, from solar and 'ground' interaction as well as internal
friction.

As our interest is in global weather patterns, we will consider the problem
on the spherical geometry, with non-negative planetary angular velocity $%
\Omega \geq 0$, and relative stream function $\psi $ - a characteristic of
2-d flows; $q$ describes the total vorticity on the sphere, 
\[
q=\omega +2\Omega \cos {(\theta )} 
\]
where the first term $\omega $ is the vorticity relative to the rotating
frame in which the sphere is at rest, and the second gives planetary
vorticity due to the angular velocity of the sphere in terms of the
rotational frequency $\Omega $ and the co-latitude $\theta $. We note the
fact that a distinguished rotating frame with fixed rotation rate $\Omega
\geq 0$ can be taken to be that in which the sphere is at rest only if the
assumption of infinite mass is made. For otherwise, the sphere will rotate
at variable rate in response to the instantaneous amount of angular momentum
in the fluid component of the coupled system, to conserve the combined
angular momentum.

Recall that the vorticity is given by the Laplacian of the stream function, 
\[
\omega =\Delta \psi 
\]
In order to study the problem in a more tractable framework, we restate the
problem in terms of local vorticity or lattice spins, following Lim \cite
{lim05}. The relative zonal and meridional flow is given by the gradient of
the stream function and will be denoted as $u_r$ and $v_r$ respectively. In
addition the flow due to rotation also contributes to the kinetic energy, it
is entirely zonal and will be denoted as $u_p$. The kinetic energy in the
inertial frame will be the key objective functional in our work, 
\begin{eqnarray}
H^{\prime }[q] &=&\frac 12\int_{S^2}dx\left[ (u_r+u_p)^2+v_r^2\right] \\
&=&\frac 12\int_{S^2}dx\left[ u_r^2+v_r^2+2u_ru_p+u_p^2\right] \\
&=&-\frac 12\int_{S^2}dxq\psi +\frac 12\int_{S^2}dxu_p^2.
\end{eqnarray}
The constant term $\frac 12\left\| u_p\right\| _2^2$ will be discarded below
- we thus work with the pseudo-kinetic energy.

Note that Stokes theorem implies that circulation on the sphere with respect
to the rotating frame is zero, 
\begin{eqnarray}
\int_{S^2}\omega = 0  \label{ST}
\end{eqnarray}

\section{The Discrete Model}

In the above discussion the relative vorticity is a function on $S^2$ given
by the Laplacian of the stream function. \cite{limnebusbook}, \cite{mean}, 
\cite{lim05}, \cite{Dinglim2} utilize a transformation to restate the
coupled fluid-sphere model in terms of a relative vorticity distribution on
the sphere which is then approximated over a discrete lattice. Proceeding
from this transformation we consider a model defined by a uniform
distribution of nodes $\{x_i\}$ on the sphere; edges spanning a pair of
sites $x_i,x_j$ are assigned interaction energy $J_{(i,j)}$. The kinetic
energy of the fluid component is then given by pairwise coupling of local
vorticity and coupling of local vorticity to the planetary spin. Without
loss of generality - for a single step renormalization based on block spins
later relates this Ising model to a more physical model with real valued
spins - we consider only local vorticity with spin of $+1$ or $-1$. As these
spins are intended to model fluid flow, Stokes' theorem requires for $2N$
vortices of equal magnitude on the sphere that $N$ vortices be of $+1$ spin
and $N$ vortices be of $-1$ spin. Under this restriction we consider the
state space with the objective of finding the statistically preferred states.

Formally, the state space consists of functions of the form 
\[
\omega (x)=\sum_iH_i(x)s_i 
\]
where 
\[
H_i=\frac{{4\pi }}N\delta (x_i-x). 
\]
The vorticity $s_i$ is assigned values $\pm 1$ which models the direction of
the spin at a node. Thus we have the following simple form for the kinetic
energy of the system, as a discrete approximation to the pseudo-kinetic
energy, 
\begin{eqnarray}
H_N=-\sum _{(i,j)}\,^{\prime }J_{(i,j)}s_is_j-\Omega \sum_iF_is_i
\label{hamiltonian}
\end{eqnarray}
where $\sum^{\prime }$ represents a summation with diagonal terms excluded, 
\begin{eqnarray}
J_{(i,j)} &=&\frac 12\int_{S^2}dxH_i(x)G(H_j)(x)  \label{int} \\
F_i &=&\int_{S^2}dx\cos \theta (x)G(H_i)(x)
\end{eqnarray}
and $G(\cdot )$ is the inverse of the Laplace-Beltrami operator on the
sphere.

The usefulness of this form is clear; it closely resembles the lattice
models of solid state physics, and becomes accessible to the rich theory of
methods of that field. In the intensive or continuum limit we see that the
coordination number of each vortex diverges. The model is not long-ranged in
the technical sense, however, as the sum of interaction energies is finite.
This differs from the Ising model in that every site is connected to every
other site in the graph, the sum being over all lattice site pairs $%
(x,x^{\prime })$. Note that as $\Omega \rightarrow 0$ the spin lattice model
becomes the pseudo-kinetic energy for the non-rotating regime. It is easily
seen that,
\begin{eqnarray}
J_{(i,j)} &=&\frac{16\pi ^2}{N^2}\ln {\left| 1-x_i\cdot x_j\right| } \\
F_i &=&-\frac{2\pi }N\cos \theta (x_i).
\end{eqnarray}

\section{Bragg-Williams Approximation}

The central feature of the Bragg method is the approximation of internal
energy of a state by its long-range order \cite{huang}. We will have to
alter the previous Bragg method, as edges in the model have different
energies as a function of their length, and furthermore, the rotation of the
planet adds a kinetic energy term analogous to an inhomogeneous magnetic
field. The method presented in this paper estimates internal energy from
local order over domains on the sphere. The implicit assumption in this
method, as in the original Bragg method is that the distribution of types of
edges is homogeneous between domains. Specifically this is done by defining
a partition of the sphere, into domains or blocks labeled $\left\{ \xi
\right\} $ which contains in principle many original lattice sites. For each
domain $\xi $ we define notation $N_{\xi} ^{+}(N_{\xi} ^{-})\equiv $ the
number of sites in cell $\xi $ which are positive(negative). Note $N_{\xi}
=N_{\xi} ^{+}+N_\xi ^{-}$. We define for every partition element $\xi $ the
local order parameter $\sigma _{\xi} $ as: 
\begin{eqnarray}
\sigma _{\xi} =2\frac{N_{\xi} ^{+}}{N_{\xi} }-1
\end{eqnarray}
The method consists of approximation of important quantities by using
-instead of the original discrete spin values - a local probability of spin
value based on the local order parameter defined by the above block
averaging procedure. Then the probability of any spin $s$ in domain $\xi $
being positive is 
\begin{equation}
P_{\xi} ^{+}\equiv Prob\left\{ s=+1|s\in \xi \right\} =\frac{1+\sigma _{\xi} 
}2,  \label{prob1}
\end{equation}
and the probability of the spin being negative is 
\begin{equation}
P_{\xi} ^{-}\equiv Prob\left\{ s=-1|s\in \xi \right\} =\frac{1-\sigma _{\xi} 
}2.  \label{prob2}
\end{equation}
The energy is the coupling of spin domains, and entropy is calculated via
the Shannon information entropy of spin variables $\left\{ s_i\right\} $.

\subsection{Statement of Equations}

The notion developed above of local order leads to a simple method of
quantifying interaction types. Specifically, pairwise interaction is
dependent on several parameters; the above interpretation lead to
derivations of equations relevant to the renormalized spin-lattice model.
Pairwise interaction occurs between all sites in three types $(ss^{\prime
})=(++)$, $(--)$ and $(+-)$. Depending on the respective domains of the
interacting sites, we use probabilities of spin distribution -
coarse-grained variables - to inform probabilities of spin interaction.
Below the probability $P_{\xi \xi ^{\prime }}^{+-}$ $=Prob\left\{
s_i=+1,s_j=-1|x_i\in {\xi },x_j\in {\xi ^{\prime }}\right\} ,$ of an edge
spanning domains $\xi $ and $\xi ^{\prime }$ being of type $(+-)$, is
calculated. Clearly, 
\begin{eqnarray}
P_{\xi \xi ^{\prime }}^{++}+P_{\xi \xi ^{\prime }}^{--}+P_{\xi \xi ^{\prime
}}^{+-}=1
\end{eqnarray}
which is identical to the relation 
\begin{eqnarray}
P_{\xi \xi ^{\prime }}^{++}+P_{\xi \xi ^{\prime }}^{--}-P_{\xi \xi ^{\prime
}}^{+-}=1-2P_{\xi \xi ^{\prime }}^{+-}  \label{PE1}
\end{eqnarray}

The LHS of(\ref{PE1}) arises below in (\ref{Bapp}), and is the sole
contribution of pairwise order to the free energy. Thus we are content to
calculate the order probability $P_{\xi \xi ^{\prime }}^{+-}$ only, as
follows $\left( \xi ^{\prime }\neq \xi \right) $: 
\[
Prob_{\xi \xi ^{\prime }}\left\{ +-\right\} =\frac{N_{\xi \xi ^{\prime
}}^{+-}}{N_{\xi} N_{\xi ^{\prime }}}=\frac{N_{\xi} ^{+}}{N_{\xi} }\frac{%
N_{\xi ^{\prime }}^{-}}{N_{\xi ^{\prime }}}+\frac{N_{\xi} ^{-}}{N_{\xi} }%
\frac{N_{\xi ^{\prime }}^{+}}{N_{\xi ^{\prime }}} 
\]
\[
=\frac{1+\sigma _{\xi} }2\frac{1-\sigma _{\xi ^{\prime }}}2+\frac{1+\sigma
_{\xi ^{\prime }}}2\frac{1-\sigma _{\xi} }2=\frac{1-\sigma _{\xi} \sigma
_{\xi ^{\prime }}}2 
\]
Analogously we have, 
\[
Prob_{\xi \xi }\left\{ +-\right\} =\frac{N_{\xi \xi }^{+-}}{\frac 12N_{\xi}
(N_{\xi} -1)}\approx \frac{N_{\xi \xi }^{+-}}{\frac 12N_{\xi} ^2}=2\frac{%
N_{\xi} ^{+}}{N_{\xi} }\frac{N_{\xi} ^{-}}{N_{\xi} } 
\]
\[
=2\frac{1+\sigma _{\xi} }2\frac{1-\sigma _{\xi} }2=\frac{1-\sigma _{\xi} ^2}%
2 
\]
for $N_{\xi} $ sufficiently large.

Finally, we state the relation 
\[
1-2P_{\xi\xi^{\prime}}^{+-}=\sigma_{\xi}\sigma_{\xi^{\prime}} 
\]
We have made this estimation based on the assumption that the number of
spins in any domain $\xi$ is sufficiently large, in fact in the
thermodynamic limit of lattice models it is standard to take the number of
spins to infinity. This must be done, however, on the finite surface of the
sphere.

\subsection{Estimation of Important Quantities}

In this section we estimate quantities in terms of the Bragg course-grained
order for the system. However, first we will note that in the process of
coarse-graining the spin values, enstrophy is now only bounded above, $%
\sigma _\xi ^2\leq 1$. Although this diverges from other approaches where
the enstrophy is constrained microcanonically, from our knowledge of the
coupled barotropic fluid-sphere physical system - otherwise known as the
Principle of Selective Decay for 2d flows - any bound on the relative
enstrophy will be sufficient to make the resulting statistical mechanics
model a well-defined one. A paper \cite{mean} by the first author
complementing this one uses a simple mean field method while enforcing
enstrophy constraints in an averaged manner.

\paragraph{Internal Energy}

The discrete approximation to the pseudo-energy (\ref{hamiltonian}) can be
rewritten in the form, 
\begin{equation}
H_N =-\frac{1}{2}\sum_{k, l} \sum_{(i,j)|(x_i,x_j) \in \xi_k \times
\xi_l}J_{(i,j)}s_is_j - \Omega \sum_k \sum_{i|x_i\in\xi_k}F_is_i
\end{equation}
where edges are summed as members of domain pairs. The factor of $\frac{1}{2}
$ arises from double counting of domains.

As consequences of the above assumptions - the distributions of positive and
negative sites is homogeneous in a partition element - we calculate the
Bragg internal energy in terms of the vector $\left\{ \sigma _\xi \right\} $%
, 
\[
H_N^B=-\frac 12\sum_{k,l}\left\langle \sum_{(i,j)|(x_i,x_j)\in \xi _j\times
\xi _k}J_{(i,j)}s_is_j\right\rangle _B-\Omega \sum_k\left\langle
\sum_{i|x_i\in \xi _k}F_is_i\right\rangle _B 
\]
where $\left\langle {}\right\rangle _B$ denotes Bragg averaging which will
be made clear below. In view of the assumption of uniformity or homegeneity,
we define the mean energy of an edge connecting two domains (or an edge from
a domain to itself) in terms of area average 
\begin{equation}
K_{\xi \xi ^{\prime }}=\left\langle \ln (1-x\cdot x^{\prime })|(x,x^{\prime
})\in \xi \times \xi ^{\prime }\right\rangle .  \label{Kdef}
\end{equation}
Further, define 
\begin{equation}
L_{\xi} =-\int_{\xi} \frac{dx}{2V_{\xi} }cos\theta (x)  \label{Ldef}
\end{equation}
to be the area-weighted, mean coupling of a site with the external field.
Returning to the energy functional, we focus on the individual terms, 
\begin{eqnarray}
\lefteqn{\left\langle \sum_{(i,j)|(x_i,x_j)\in \xi \times \xi \prime
}J_{(i,j)}s_is_j\right\rangle _B} \\
&=&\left\langle \sum_{(i,j)|(x_i,x_j)\in \xi \times \xi ^{\prime
}}J_{(i,j)}\delta \left( 1-s_is_j\right) -\sum_{(i,j)|(x_i,x_j)\in \xi
\times \xi ^{\prime }}J_{(i,j)}\delta \left( 1+s_is_j\right) \right\rangle _B
\\
&=&\left( P_{\xi \xi ^{\prime }}^{++}+P_{\xi \xi ^{\prime }}^{--}-P_{\xi \xi
^{\prime }}^{+-}\right) \left\langle \sum_{(i,j)|(x_i,x_j)\in \xi \times \xi
^{\prime }}J_{(i,j)}\right\rangle  \label{Bapp} \\
&=&\sigma _{\xi} \sigma _{\xi ^{\prime }}V_{\xi} V_{\xi ^{\prime }}K_{\xi
\xi ^{\prime }}
\end{eqnarray}
In (\ref{Bapp}) we have the course-graining approximation over the edges in $%
\xi \times \xi ^{\prime }$. The expected interaction of a partition element
with itself differs by a factor of $\frac 12$ to account for double
counting, 
\begin{eqnarray}
\lefteqn{\left\langle \sum_{(i,j)|(x_i,x_j)\in \xi \times \xi
}J_{(i,j)}s_is_j\right\rangle } \\
&=&\frac 12\left\langle \sum_{(i,j)|(x_i,x_j)\in \xi \times \xi
}J_{(i,j)}\delta \left( 1-s_is_j\right) -\sum_{(i,j)|(x_i,x_j)\in \xi \times
\xi }J_{(i,j)}\delta \left( 1+s_is_j\right) \right\rangle _B \\
&=&\frac 12\left( P_{\xi \xi }^{++}+P_{\xi \xi }^{--}-P_{\xi \xi
}^{+-}\right) \left\langle \sum_{(i,j)|(x_i,x_j)\in \xi \times \xi
}J_{(i,j)}\right\rangle  \label{Bapp2} \\
&=&\frac 12\sigma _{\xi} ^2V_{\xi} ^2K_{\xi \xi }
\end{eqnarray}
Finally, the external interaction becomes 
\[
\left\langle \sum_{i|x_i\in \xi }F_is_i\right\rangle _B=N_\xi \text{ }%
\left\langle s_i|x_i\in \xi \right\rangle \text{ }\left\langle F_i|x_i\in
\xi \right\rangle =-\sigma _{\xi} \int_{\xi} dx\text{ }\frac 12cos\theta
(x)=\sigma _{\xi} V_{\xi} L_{\xi} 
\]
Thus we have the simple expression for the Bragg internal energy 
\begin{equation}
U=-\frac 12\sum_{i,j}\sigma _{\xi _i}\sigma _{\xi _j}V_{\xi _i}V_{\xi
_j}K_{\xi _i\xi _j}-\Omega \sum_{i}\sigma _{\xi _i}V_{\xi _i}L_{\xi _i}.
\label{bragu}
\end{equation}

\paragraph{Entropy}

It should be noted that the nature of our model is constrained to a constant
sized sphere, and therefore no extensive quantities exist, only intensive
ones. Since the coarse-grained or renormalized spins $\left\{ \sigma _\xi
\right\} $ are equivalent to the probability distributions in (\ref{prob1})
and (\ref{prob2}), we therefore calculate -as is standard in equilibrium
statistical mechanics- the Shannon entropy of the renormalized or
coarse-grained state in terms of the local order parameters $\left\{ \sigma
_\xi \right\} $,

\begin{equation}
S=-k_B\sum_iV_{\xi _i}\left[ \frac{1+\sigma _{\xi _i}}2\ln \frac{1+\sigma
_{\xi _i}}2+\frac{1-\sigma _{\xi _i}}2\ln \frac{1-\sigma _{\xi _i}}2\right] .
\label{entrob}
\end{equation}
This entropy is weighted by area of the domains.

\paragraph{The Free Energy}

The Helmholtz free energy is then, 
\[
\Psi =U-TS 
\]
\[
=-\frac 12\sum_{i,j}\sigma _{\xi _i}\sigma _{\xi _j}V_{\xi _i}V_{\xi
_j}K_{\xi _i\xi _j}-\Omega \sum_i\sigma _{\xi _i}V_{\xi _i}L_{\xi _i} 
\]
\[
+Tk_B\sum_iV_{\xi _i}\left[ \frac{1+\sigma _{\xi _i}}2\ln \frac{1+\sigma
_{\xi _i}}2+\frac{1-\sigma _{\xi _i}}2\ln \frac{1-\sigma _{\xi _i}}2\right] 
\]
constrained to the set of $\sigma _\xi $ which are solutions to 
\begin{equation}
0=\sum_iV_{\xi _i}\sigma _{\xi _i} .  \label{Ce}
\end{equation}
The goal is to find critical points of the free energy with respect to $%
\left\{ \sigma _{\xi _i}\right\} _i$. Enforcing $(\ref{Ce})$, we get a
Lagrange multiplier problem. We must find the critical points, given by the
simultaneous solution of the m equations, 
\[
\lambda V_{\xi _i}\left\{ \nabla _{\sigma _{\xi _i}}\sum_l\sigma _{\xi
_l}\right\} =\left\{ \nabla _{\sigma _{\xi _i}}\Psi \right\} =-V_{\xi
_i}\Omega L_{\xi _i}-\sum_l\sigma _{\xi _i}V_{\xi _i}V_{\xi _l}K_{\xi _i\xi
_i}+Tk_BV_{\xi _i}\left( \frac 12\ln \frac{1+\sigma _{\xi _i}}{1-\sigma
_{\xi _i}}\right) 
\]
along with the constraint equation $(\ref{Ce})$. Equivalently we write, $%
\forall i=1,..,m,$%
\[
\sigma _{\xi _i}=\tanh \left[ \beta (\Omega L_{\xi _i}+\lambda )+\beta
\sum_l\sigma _{\xi _l}V_{\xi _l}K_{\xi _i\xi _l}\right] 
\]
which is a $m-$ dimensional fixed point equation for $\{\sigma _{\xi
_i}\}_{i=1}^m.$ We will show that a well-defined continuum limit for this
equation exists in the form a fixed point equation in Hilbert space $%
L_2(S^2).$

\section{Polar State Criteria}

In this section we implement the Bragg method on a two domain partition of
the sphere. We find the order of the system varies continuously with the
temperature and planetary spin. The system exhibits a continuous phase
transition of second order at a negative critical temperature in the
non-rotating case. The large spin $\Omega >0$ thermodynamic regimes exhibit
a continuous phase transition at a positive critical temperature between a
weakly counter-rotating ordered state and a mixed state. A contionuous transition
at $\beta =0$ near-mixed counter-rotating states at high positive temperatures 
and near-mixed pro-rotating states at negative temperatures far from zero. 
A final continuous phase transition in the negative regime occurs when 
negative temperatures approaches zero to an ordered strongly pro-rotating state.

The sphere is partitioned into the northern and southern hemispheres,
labeled $1$ and $2$ respectively. In this case we have only parameters 
$\sigma _1$ and $\sigma _2$ representing local order parameters of the 
northern and southern hemispheres respectively. Then $\sigma _2=-\sigma _1$ 
from $(\ref{ST})$. This leads a simple approximation of the free energy 
\begin{equation}
\Psi = \sigma_1^2 V_1^2 \left( K_{1,2} - K_{1,1}\right) + 2 \Omega \sigma  
+ Tk_B 2\pi \left[ \left( 1+\sigma_1\right)\ln{\frac{1+\sigma_1}{2}}
+ \left( 1-\sigma_1\right)\ln{\frac{1-\sigma_1}{2}}\right]
\label{FE2domain}
\end{equation}
and fixed point equations for the extremal values
\[
\sigma _1=\tanh {\left[ \beta (\Omega L_1+\lambda )+\beta \sigma
_1V_1(K_{1,1}-K_{1,2})\right] } 
\]
\[
-\sigma _1=\tanh {\left[ -\beta (\Omega L_1-\lambda )-\beta \sigma
_1V_1(K_{1,1}-K_{1,2})\right] ,} 
\]
which reduces easily to the fixed point equation in one variable 
\begin{equation}
\sigma _1=\tanh {\left[ \beta \Omega L_1+\beta \sigma
_1V_1(K_{1,1}-K_{1,2})\right] .}  \label{poleq}
\end{equation}
Here $L_1 = -1$ and $V_1 = 2\pi$. 
A simple monte carlo integrator was used to 
estimate $K = -V_1^2\left( K_{1,1} - K_{1,2}\right) \approx 120$.

\paragraph{Non-rotating}

The non-rotating case is given in equation (\ref{poleq}) by setting $\Omega
=0$. For positive $\beta $ the RHS of (\ref{poleq}) is decreasing while the
LHS increases, thus there exists a unique solution $\sigma _1=0,$ which
corresponds to the mixed state.

\begin{figure}[tbh]
\begin{center}
\includegraphics{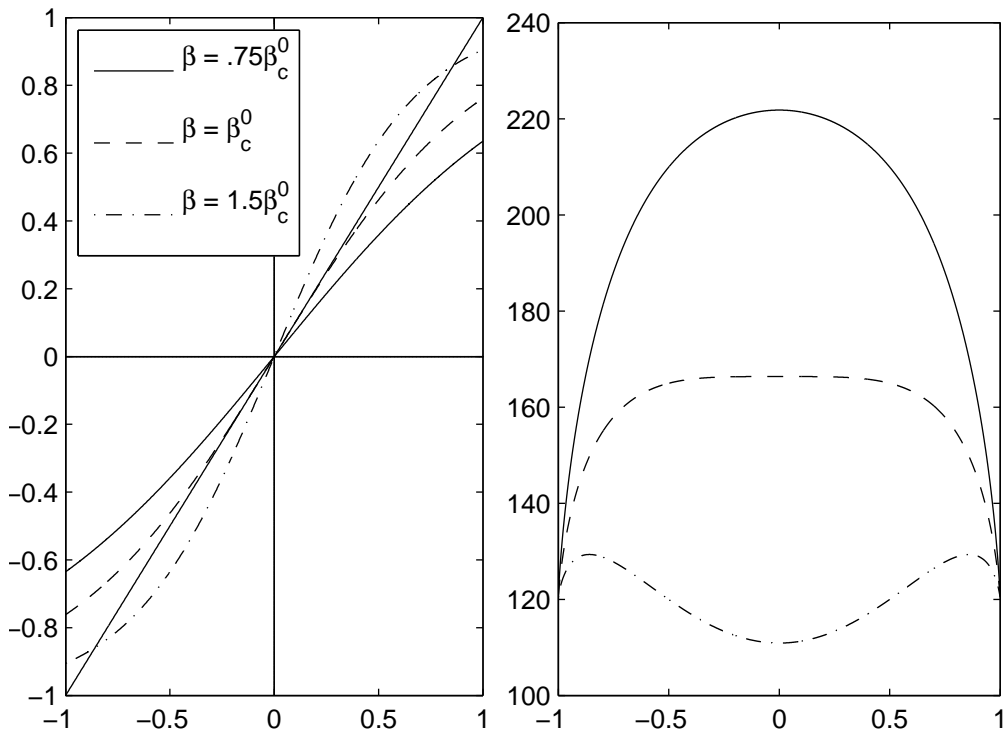}
\end{center}
\caption{Left: The LHS and RHS for several values of $\beta$ of equation $(\ref{poleq})$.
Right: Corresponding plots of the free energy. For all plots $\Omega =0$.}
\label{f2}
\end{figure}

In the negative temperature domain $\beta <0$ we must maximize the free
energy, according to an easy extension of Planck's theorem to negative
temperatures. For this case it is clear that a solution is found at $\sigma
_1=0$, corresponding to a mixed state. Whether other fixed points exist
depends on the slope of the RHS of $(\ref{poleq})$ at $\sigma _1=0$.
Graphically it is clear this situation arises when the slope of the RHS has
a slope of one, thus there is a critical quantity $\beta _c^0$ given by $%
\beta _c^0V_1(K_{11}-K_{12})=1$. Thus for 
\[
0>\beta >\beta _c^0= - \frac{2 \pi}{K}
\]
the point $\sigma _1=0$ is the only stationary point(see figure $(\ref{f2})$%
).

For $\beta <\beta _c^0<0$ there exist non-zero maximizers of the free energy $%
\sigma ^{-}<0$, $\sigma ^{+}>0$; the tanh function is odd, thus $\sigma
^{-}=-\sigma ^{+}$(figure (\ref{f2})). 
%\begin{figure}[tbh]
%\begin{center}
%\includegraphics{braggfig3}
%\end{center}
%\caption{A graph of the fixed point when $\beta <\beta _c<0$,$\Omega =0$.}
%\label{f3}
%\end{figure}
The free energy is symmetric here so we need only consider fixed points $%
\sigma ^{+}$ and $\sigma =0$.
Consider the Taylor series of the free energy around zero
\begin{eqnarray}
0 &<&\Phi(\sigma ) - \Phi(0) \\
\mbox{} &=&(K + 2\pi Tk_B)\sigma ^2 + \frac{4\pi}{3}Tk_B\sigma^4 + O\left( \sigma^6\right)
\end{eqnarray}
We therefore have a polarized ordered state for $\beta <\beta _c^0<0,$ which
corresponds to solid-body rotating flow. A standard symmetry-breaking phase
transition is thus predicted by the fixed point equation in this Bragg mean
field model, in agreement with both the simple mean field theory \cite{mean}
and Lim's spherical model \cite{Limsphere06}, \cite{iutam06}, \cite{limnebus}%
, \cite{Dinglim}, \cite{Dinglim2}. \\  We have established the result,

\smallskip\ 

\textbf{Proposition 1: }\textit{In the non-rotating case of the Bragg model,
there is a negative temperature phase transition to a solid-body-rotating
ordered state for }$\beta <\beta _c^0=-2\pi/K<0.$\textit{\
For all other values of the temperature - positive and negative - the most
probable state is a mixed vorticity state.}

\smallskip\ 

\paragraph{Rotating}

In the rotating regime the free energy can be approximated by a series as
\begin{equation}
\Psi_{T,\Omega} \left( \sigma\right) = -4\pi Tk_B \ln{2} + 2\Omega\sigma 
+ \left( K + 2\pi Tk_B \right) \sigma^2 + \frac{\pi Tk_B}{3}\sigma^4 + O\left[ \sigma^6\right] 
\label{FE2Drot}
\end{equation}

For $\Omega >0$, and independent of whether the temperature is positive or
negative, the RHS of the fixed point equation $(\ref{poleq})$ is zero at
\begin{equation}
\bar{\sigma}_1=\frac{-2\pi\Omega}{K}<0,  \label{posrot}
\end{equation}
since $K>0.$ Moreover, it is easy to see that the zero $\bar{\sigma}_1$ satisfies 
\begin{equation}
-1<\bar{\sigma}_1<0  \label{bee1}
\end{equation}
if and only if planetary spin $\Omega $ - taken to be non-negative by
convention - satisfies the condition 
\begin{equation}
0<\Omega <\Omega _{cc}=\frac{K}{2\pi}  \label{baa1}
\end{equation}
where $\Omega _{cc}>0$ is independent of temperature - that is, planetary
spins are not too large. This condition turns out to be significant for the
determination of the fixed points of $(\ref{poleq})$ in the rotating case
which separates naturally into positive and negative temperature subcases
considered next.

\subsubsection{Positive temperatures}

In the case of positive temperatures, by virtue of the property that the RHS
of $(\ref{poleq})$ is decreasing in $\sigma _1$ and the fact that it is
bounded between $\pm 1,$ there is always a fixed point $\sigma _1$ of $(\ref
{poleq})$ satisfying 
\[
-1<\sigma _1<0.
\]
By virtue of the shape of the graph of RHS of $(\ref{poleq}),$ this fixed
point also satisfies 
\begin{equation}
\bar{\sigma}_1<\sigma _1<0.  \label{boo1}
\end{equation}
Furthermore, as $\beta \rightarrow \infty$, $\sigma_1 \searrow \max{\left( \bar{\sigma}_1,-1\right) } $.
Since the zero $\bar{\sigma}_1$ satisfies 
\begin{equation}
-1/2<\bar{\sigma}_1<0  \label{boo2}
\end{equation}
if the planetary spin $\Omega $ is smaller than the threshold value 
\begin{equation}
\Omega _c=\frac{K}{4\pi}>0,  \label{bee2}
\end{equation}
we deduce from equation (\ref{boo1}) that the fixed point also satisfies 
\begin{equation}
-1/2<\sigma _1<0  \label{boo3}
\end{equation}
for planetary spins smaller than $\Omega _c.$ From the fixed point 
equation(\ref{poleq}) we find a phase transition where $\sigma < -1/2$
at positive critical temperature
\[T^+\left( \Omega\right) k_B = \frac{4\pi\Omega-K}{2\pi\ln{3}}\]
Thus, because $\Omega _c$ is independent of the inverse temperature $\beta > 0,$ 
we have established the following result where we have taken the cutoff $|\sigma _1|=1/2$ to
separate mixed states from ordered states,

\smallskip\ 

\textbf{Proposition 2: }\textit{\ (a) The most probable state in the Bragg
model is the mixed vorticity state for all positive temperatures when
planetary spins are smaller than }$\Omega _c$\textit{\ -  there are
therefore no phase transitions in positive temperatures in this case. (This
is shown in figure (\ref{TSVsm}) where the fixed points are plotted vs }$T$\textit{\
for very small }$\Omega <\Omega _c.)$ (b) \textit{On the other hand for
planetary spins that are not small, that is }
\begin{equation}
\Omega \geq \Omega _c=\frac{K}{4\pi},  \label{bee3}
\end{equation}
\textit{the fixed point }$\sigma _1(\beta )\in (-1,0)$\textit{\ has
continuously increasing long range order as temperature decreases (see
figure }$\ref{TSVlg}$\textit{\ below) and below}$T_c^{+}(\Omega )>0,$%
\textit{\ the statistical equilibria }$\sigma _1<-1/2$\textit{\ is an
organized counter-rotating physical flow. }

\smallskip\ 

\begin{figure}[th]
\begin{center}
\includegraphics{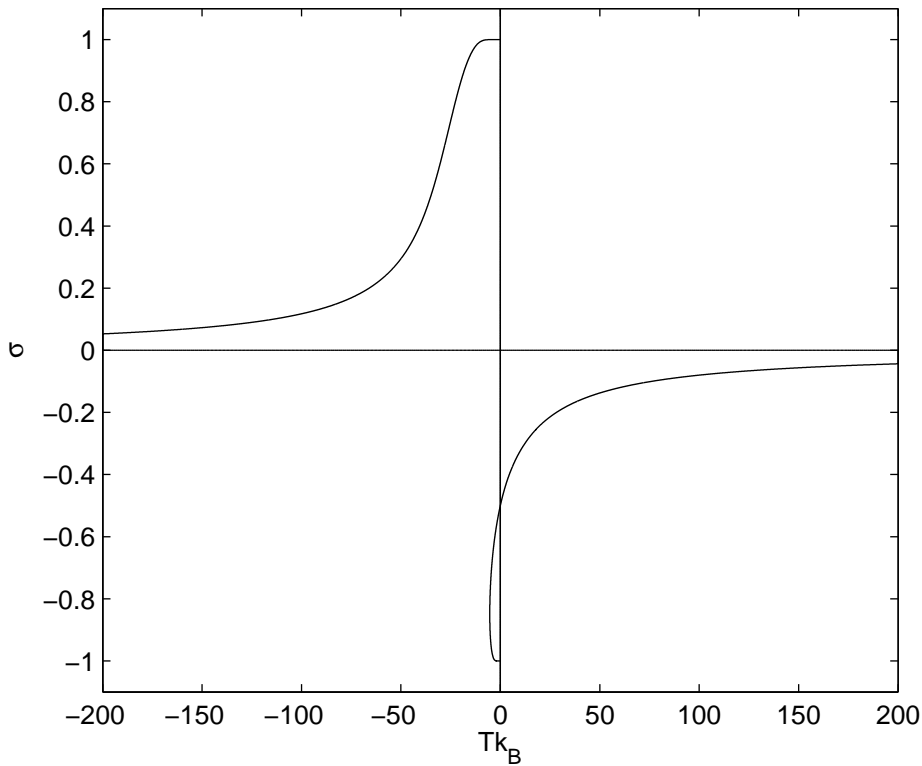}
\end{center}
\caption{A graph of $\sigma$ vs $Tk_B$ when $\Omega  = \Omega_c$.}
\label{TSV}
\end{figure}

This is in agreement with the result obtained in the simple mean field
approach in the case of positive temperatures \cite{mean}. Properties of the
first threshold value of planetary spin are clearly
shown in figure (\ref{TSV}) and figure 3 where fixed points and free energy respectively
are plotted vs $T$ for
planetary spin $\Omega = \Omega_c $.

%\begin{figure}[th]
%\begin{center}
%\includegraphics{TvsSigmaVis3Vcon2}
%\end{center}
%\caption{A graph of $Tk_B$ vs $\sigma$ when $\Omega  = \frac{3}{2}\Omega_c$.}
%\label{TSA}
%\end{figure}

\begin{figure}[th]
\begin{center}
\includegraphics{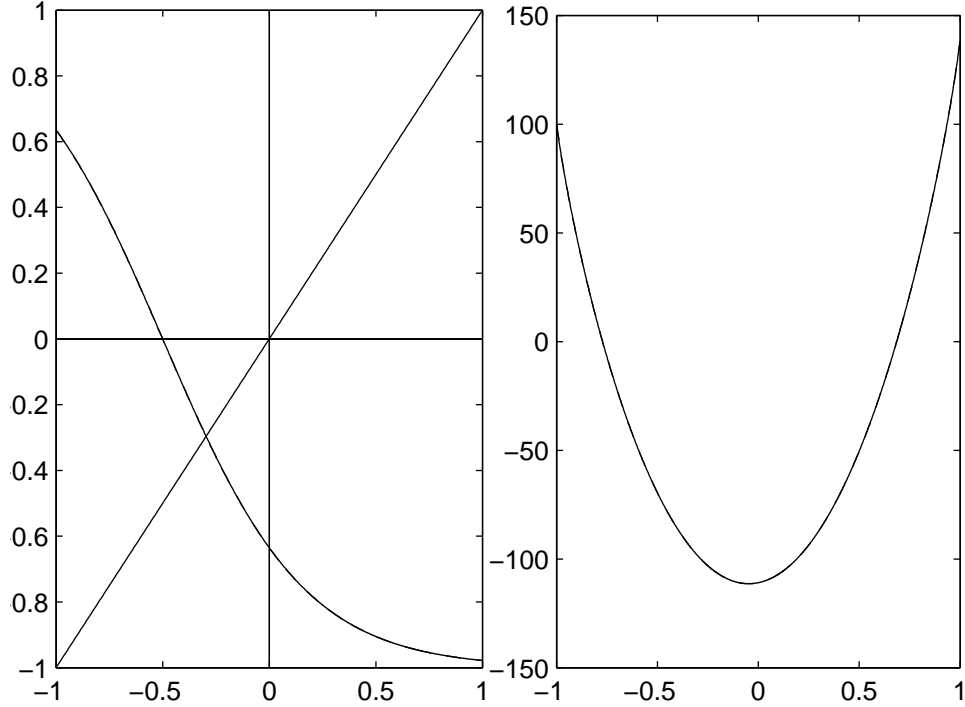}
\end{center}
\caption{LEFT: A graph of the fixed point when $\beta > 0,\Omega = \Omega_c$.
RIGHT: The free energy}
\label{f1}
\end{figure}

\subsubsection{Transition between positive and negative temperatures with
large $|T|$}

For all values of planetary spins $\Omega ,$ the most probable state changes
smoothly through mixed states between high positive temperatures $T\gg 1$
and large-absolute-valued negative temperatures $T\ll -1$. At $T\gg 1,$ the
preferred mixed state or fixed point has a slight negative angular momentum
- it is slightly anti-rotating. For $T\ll -1,$ it is the reverse - the fixed
point is a mixed state with a slight positive angular momentum or
pro-rotation bias. These facts are clearly shown in
figures (\ref{TSVlg}) and (\ref{TSVsm}) for several distinct values of $\Omega .$

\subsubsection{Negative temperatures}

We divide the analysis for negative temperatures into two natural categories
- where the fixed point equation has the possibility of multiple fixed
points and where it always have exactly one fixed point. In both categories,
as negative $T$ increases (or decreases in absolute value), and for all values of planetary spin,
the slightly
pro-rotating mixed state described in the last subsection gives way
continuously to a strongly pro-rotating state at $\beta ^{-}(\Omega )<0$ -
this is chosen to be the value of $\beta $ for which the fixed point 
$\sigma_1(\beta ^{-})=0.5.$ This is clearly depicted in figures 2, 4 and 5 for 
several values of the planetary spin.

Solving (\ref{poleq}) for $Tk_B$ gives
\begin{equation}
Tk_B = \frac{2\pi\Omega+\sigma K }{\pi\ln \left( \frac{1-\sigma}{1+\sigma} \right) }.
\end{equation}
Thus,
\begin{equation}
\beta^-(\Omega) = \frac{-2\pi\ln{3}}{4\pi\Omega + K}
\label{betnV}
\end{equation}

\begin{figure}[th]
\begin{center}
\includegraphics{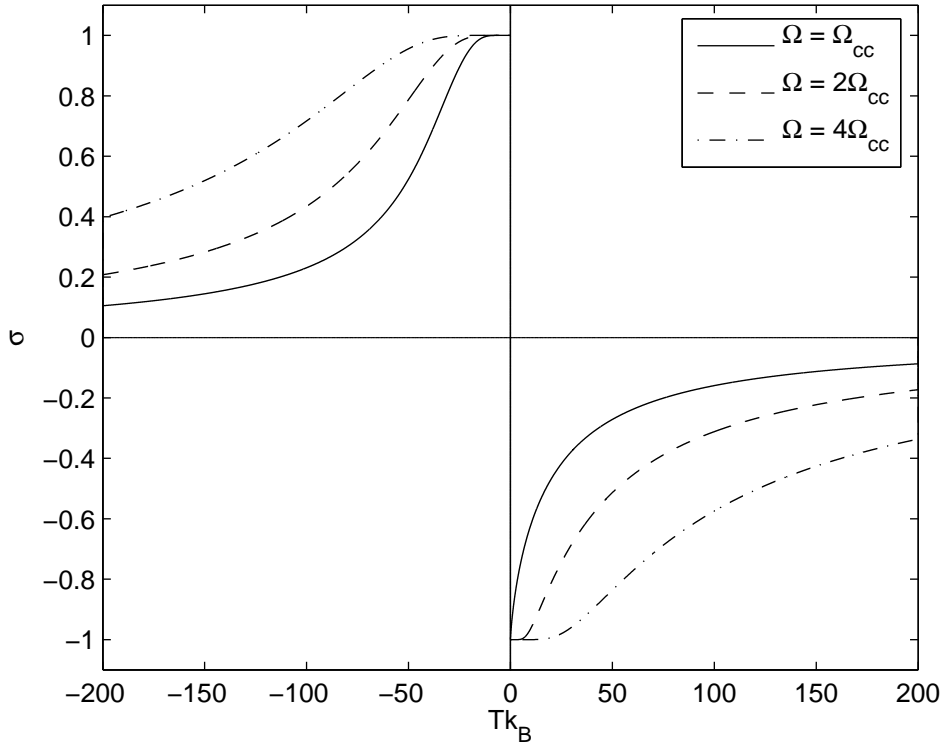}
\end{center}
\caption{A graph of $\sigma$ vs $Tk_B$ when $\Omega  > \Omega_{cc}$}
\label{TSVlg}
\end{figure}

\begin{figure}[tbh]
\begin{center}
\includegraphics{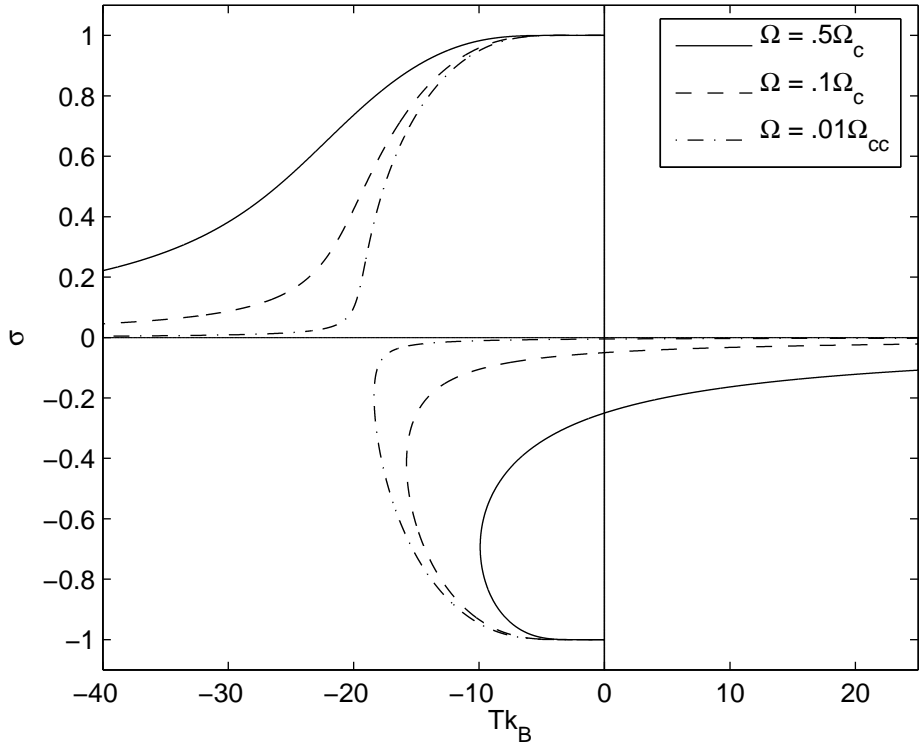}
\end{center}
\caption{A graph of $\sigma$ vs $Tk_B$ when $\Omega_c>\Omega >0$.}
\label{TSVsm}
\end{figure}

However, figures 2, (\ref{TSVlg}) and (\ref{TSVsm}) also indicate that there is
another temperature threshold when the spin is not too large, namely 
\begin{equation}
\beta_c^\Omega < 0  \label{imp1}
\end{equation}
where $\beta _c^\Omega $ $<0$ is the value for which multiple fixed points
appear when $\beta \leq \beta _c^\Omega $ and when $\Omega <\Omega _{cc}$ -
both defined below. We can find an estimate for $\beta_c^\Omega$. 
Critical value leading to the apearance of two more counter-rotating solutions of (\ref{poleq}) is given by
\begin{equation} 
0 = \frac{dTk_B}{d\sigma} 
= \frac{K\left( 1-\sigma^2\right) \ln\left( \frac{1+\sigma}{1-\sigma}\right) + 4\pi\Omega +\sigma 2 K }
{\pi\left( 1-\sigma^2\right) \left( \ln\left( \frac{1-\sigma}{1+\sigma}\right) \right) ^2}
\label{Textremum}
\end{equation}
from which it follows 
\begin{equation}
\beta_c^\Omega = \frac{-2\pi }{K\left( 1-\sigma^2\right) }\vert_{\sigma=\sigma^\Omega}
\label{betacV}
\end{equation}
An approximation of the numerator on the RHS of \ref{Textremum} by a third order 
polynomial in $\sigma$ gives
\begin{equation}
\sigma^\Omega \approx -\left( \frac{3\pi\Omega}{K}\right)^{1/3} 
\label{sV}
\end{equation}

We can now conclude that

%We will state and prove the inequality in (\ref{imp1}) in
%****

%From \ref{betnV} and \ref{betacV}, as $\Omega \rightarrow 0$,

%\[\beta\left( \Omega\right)  = \frac{-2\pi\ln{3}}{K} < \frac{-2\pi}{K} = \beta_c^\Omega\]
%Thus,
%\[T^{0^+}_c > T (0^+) \]
%so the appearance of multiple states precedes the ordered phase.

%****
\smallskip\ 

\textbf{Lemma 3:} \textit{The critical temperature}$T_c^\Omega$ \textit{is increasing in} $\Omega$

\smallskip\ 

Proof: Follows imeadiately from (\ref{betacV}) and (\ref{sV})
%\[
%\Omega _c\leq \Omega <\Omega _{cc}
%\]
%property (\ref{imp1}) is a direct corollary of the Lemma in %(\ref{facto})
%and the above definition of $\beta ^{-}(\Omega )$ as the inverse temperature
%at which $\sigma _1(\beta ^{-})=0.5.$
%This completes the proof of Lemma 3.

%Another way to state the property in (\ref{imp1}) is the transition between
%mixed and strongly pro-rotating states occur at a cooler negative
%temperature than the temperature threshold at which multiple stationary
%states first appear if allowed.

\paragraph{One fixed point}

In the case of negative temperatures, the physical bound $|\sigma _1|\leq 1,$
in the fixed point equation implies that there is another threshold on the
planetary rotational rate. When 
\begin{equation}
\Omega \geq \Omega _{cc}=\frac{V_1(K_{11}-K_{12})}{L_1}=2\Omega _c,
\label{omegac}
\end{equation}
there are only pro-rotating solutions $\sigma _1>0$ to the fixed point
equation because any negative fixed point $\sigma _1^{-}<0$ - if they exist
- must satisfy 
\[
\sigma _1^{-}<\bar{\sigma}_1<0
\]
where the zero $\bar{\sigma}_1$ must in turn satisfy 
\[
\bar{\sigma}_1<-1
\]
in view of (\ref{bee1}) and (\ref{baa1}), which leads to the contradiction $%
\sigma _1^{-}<-1.$ This is depicted in figure (\ref{TSVlg}) for $\Omega >\Omega _{cc}.$

Thus, we have the result,

\smallskip\ 

\textbf{Proposition 4: }\textit{\ For large spins }$\Omega >\Omega _{cc},$%
\textit{\ there is a single negative- temperature transition at }$\beta
^{-}(\Omega )<0$\textit{\ between the mixed and the strongly pro-rotating
state for }$|T|\ll 1.$

%Using Lemma 3, we have the following result, which is again clearly depicted
%in figure (\ref{TSA}) for $\Omega \in (\Omega _c,\Omega _{cc}).$

%\smallskip\ 

%\textbf{Proposition 5: }\textit{\ For spins not too large, that is} \textit{%
%if }$\Omega \leq \Omega _{cc},$\textit{\ then for intermediate inverse
%temperature such that }$\beta _c^\Omega <\beta <\beta ^{-}(\Omega )<0,$%
%\textit{\ there is exactly one fixed point, namely the pro-rotating most
%probable state, }$1/2<\sigma _1<1,$\textit{\ in the Bragg model. For less
%hot negative inverse temperatures such that  }$\beta ^{-}(\Omega )<\beta <0,$
%\textit{the most probable state is a mixed state with a pro-rotating bias
%which grows as }$\beta \searrow \beta ^{-}(\Omega ).$

%\smallskip\ 

\paragraph{Multiple fixed points}

On the other hand, when $\Omega <\Omega _{cc},$ it is possible for the
argument of tanh to be zero at values of $\bar{\sigma}_1$ within the
physical range $-1<\bar{\sigma}_1<0,$ making it in turn possible to have
counter-rotating or mixed- state fixed points that satisfy 
\[
-1\leq \sigma _1<\bar{\sigma}_1<0
\]
if in addition, 
\[
\beta \leq \beta _c^\Omega <0.
\]

In the case $\Omega \leq \Omega _{cc},$ and $\beta \leq \beta _c^\Omega ,$
there are generically three fixed points due to the structure of tanh, one
pro-rotating and in general two others - one of which is strongly
counter-rotating and one mixed with a small counter-rotation - which merge
into a single degenerate counter-rotating / mixed solution when $\beta
=\beta _c^\Omega $. This last threshold $\beta _c^\Omega <0$ is a
consequence of the fact that $-\beta $ determines the slope of tanh near $0$
and a large enough slope is required in order for the graph of tanh to
intersect the line of the identity function. 
%Once again the distinct
%categories - where there is the possibility of three fixed points and where
%there is always exactly one fixed point - are shown in the set of figures \ref{TSVlg},
%, \ref{TSVsm} and (\ref{f1}) taken together.

\begin{figure}[tbh]
\begin{center}
\includegraphics{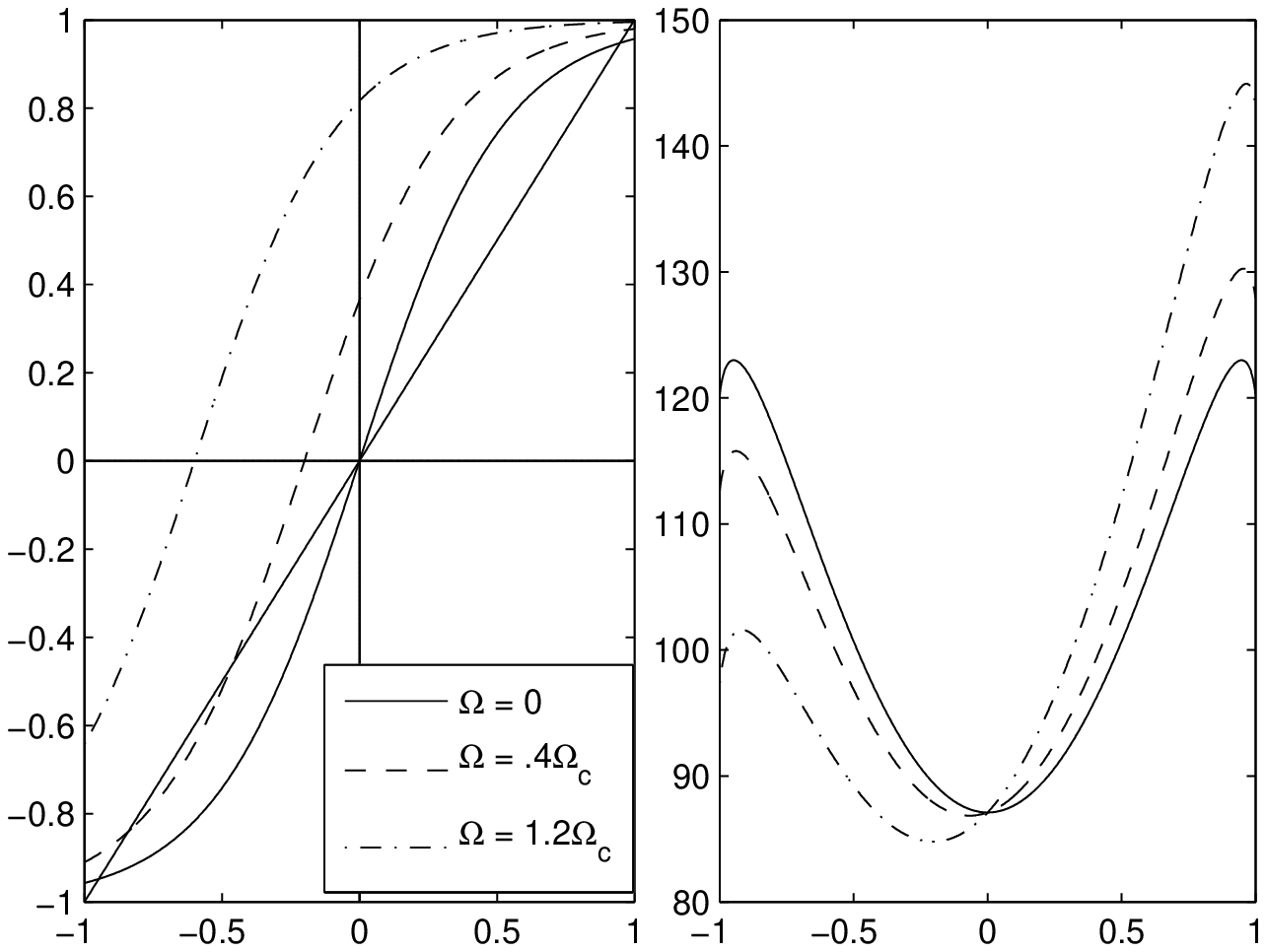}
\end{center}
\caption{LEFT:Plots of the fixed points RIGHT: The free energy. 
Both plots are for the same value of $\beta <\beta _c^\Omega <0$}
\label{f5}
\end{figure}

%In the case of multiple fixed points, maximization of the free energy at
%these three fixed points decides which is the most probable coarse-grained
%vorticity state. We will next consider the free energy of these states.
%The next result is key to proving property (\ref{imp1}).

There is a transition at $\beta ^{-}(\Omega)<0$ between the mixed and
the strongly pro-rotating state and another negative temperature
transition $\beta _c^\Omega <0$ when $\Omega < \Omega_{cc}$ where aditional 
critical points arise in (\ref{FE2Drot}).
The Bragg free energy has a simple form in this case which we can exploit 
to easily see that the pro-rotating solution has greater free energy than
the counter rotating solution in negative temperatures close to zero, as depicted in
figure 6 for
$ \beta > \beta_c^\Omega .$

In the two domain case, $\Omega > 0$ adds a linear term to the free energy functional,
\begin{equation}
\Psi_{T,\Omega}\left(\sigma\right)  = \Psi_{T,0}\left(\sigma\right) + 2\Omega\sigma
\label{FisFoplusL}
\end{equation}
Recall from the nonrotating case that for $0 > T > T^0_c$ the free energy has two 
maximizers we will denote here as $\bar{\sigma}^-$ and $\bar{\sigma}^+$; the
free energy in this case is even so $\bar{\sigma}^-= - \bar{\sigma}^+$. 
Therefore, from Lemma 3 and equation (\ref{FisFoplusL}), 
\begin{equation}
\Psi_{T,\Omega}\left( \bar{\sigma}^-\right) < \Psi_{T,\Omega}\left( \sigma^-\right) 
< \Psi_{T,\Omega}\left( \bar{\sigma}^+\right) <  \Psi_{T,\Omega}\left( \sigma^+\right)
\end{equation}
It is also of interest to note that $ \bar{\sigma}^- < \sigma^- < 0
< \bar{\sigma}^+ < \sigma^+.$
In physical terms this implies that the ordered state $\sigma _1^{+}>0$ (in
which positive relative vorticity dominates the northern hemisphere), is
`more ordered' than the symmetric solutions at $\Omega = 0$
which is in turn more ordered than the counter-rotating fixed 
point $\sigma _1^{-}<0.$ We have demonstrated

\smallskip\ 

\textbf{Proposition 5: }\textit{\ For spins not too large, that is} \textit{%
if }$\Omega \leq \Omega _{cc},$\textit{\ then for hottest inverse
temperature such that }$\beta <\beta _c^\Omega <0,$\textit{\ there are
exactly three fixed points in the Bragg model but the  most probable state
is pro-rotating such that }$\sigma _1 \nearrow 1.$

\subsection{Summary of main results}

In summary, the simple two domains case of the fixed point equation predicts
that in the rotating problem, there are two critical values of the planetary
spin, $\Omega _c$ in (\ref{bee2}) and $\Omega _{cc}=2\Omega _c.$ Below $%
\Omega _c$, there is no transition at positive temperatures and the most
probable barotropic flow state is mainly a mixed state (\ref{boo3}) with
some small amount of counter-rotation which vanishes as temperature
increases. For planetary spins above $\Omega _c,$ there is a continuous
positive temperature transition from mixed states at $T>T_c^{+}(\Omega )$ to
strongly counter-rotating barotropic states at lower positive temperatures.

For all values of spin, there is a smooth transition through mixed states at 
$\beta =0$ between slightly anti-rotating mixed states for $T\gg 1$ and
slightly pro-rotating mixed states for negative $T$ with $|T|\gg 1.$ Again
for all values of spin $\Omega $, there is a continuous transition at $\beta
^{-}(\Omega )<0$ between mixed states for cooler negative $\beta >\beta
^{-}(\Omega )$ and strongly pro-rotating states for hotter $\beta <\beta
^{-}(\Omega ).$ For large spins $\Omega >\Omega _{cc},$ this is the only
possible transition at negative temperatures.

For intermediate to small values of spin, $\Omega <\Omega _{cc},$ there is a
negative threshold $\beta _c^\Omega$ at which the possibility for multiple 
(meta-stable) thermodynamic equilibria first arises. 
We argued that when there are multiple fixed points, the
pro-rotating branch has the largest free energy and thus continues to be the
thermodynamically stable macrostate. We conclude that for all spins, there
is a single negative- temperature- transition at $\beta ^{-}(\Omega )<0$ to
the pro-rotating state for the hottest negative temperatures - those with
the smallest absolute values corresponding to the largest kinetic energy.

The complete analysis of the transitions in the non-rotating case and in the
rotating case agrees well with the simple mean field theory and Lim's
spherical model.

\section{The infinite dimensional nonextensive limit}

As noted above, the model does not have an extensive limit from which we can
derive intensive quantities. We have therefore bypassed the standard
thermodynamic construction and directly built an non-extensive model.
Convergence theorems support our continuum model. In this frame work we have
had no need to include stipulations on the types configuration of domains.
In light of this, we will consider a sequence of domain systems which are
constructed from Voronoi cells of a uniform mesh as in Lim and Nebus\cite
{limnebus}. This affords us an advantage that we may assume the coarse
grained domains to be spatially symmetric. We henceforth refer to the
sequence of domain systems and their associated ensembles as a Bragg
process. The limiting ensemble of a Bragg process is the space $L_2(S^2)$ of
functions $\sigma :S^2\mapsto \left[ -1,1\right] $, which has an expression
of the free energy analogous to the discrete case, 
\[
\Psi \left[ \sigma \right] =-\frac 12\int dx\int dy\sigma (x)\sigma
(y)K(x,y)-\Omega \int dx\sigma (x)L(x) 
\]
\[
+Tk_B\int dx\left[ \frac{1+\sigma (x)}2\ln \frac{1+\sigma (x)}2+\frac{%
1-\sigma (x)}2\ln \frac{1-\sigma (x)}2\right] \label{s2eq} 
\]
from which we can easily recover the fixed point equation. The free energy
must be extremized over the subspace of functions constrained by Stokes'
theorem $\Sigma =\left\{ \sigma :S^2\mapsto [-1,1]|0=\int_{S^2}dx\sigma
(x)\right\} $. Another physical constraint namely bounded relative enstrophy
- the square of the $L_2$ norm of the coarse-grained relative vorticity
field $\sigma (x)$ - is implicit in the condition that $|\sigma |\leq 1$ in
the fixed point equation which can be derived in the continuous case
similarly to the above discrete case, 
\begin{equation}
\sigma (x)=\tanh {\left[ \beta (\Omega L(x)+\lambda )+\beta
\int_{S^2}dy\sigma (y)K(x,y)\right] .}  \label{nleq}
\end{equation}

It is clear from the definition of $K$ (\ref{Kdef}) that we have recovered
the inverse Laplacian $G$ and we can rewrite (\ref{nleq}) as 
\begin{equation}
\Delta \psi =\tanh {\left[ \beta \left( \Omega L(x)+\lambda +\psi \right)
\right] }  \label{nleq2}
\end{equation}
where $\psi $ is the course grained stream function. From this form,
remembering that $L(x)$ is proportional to the first spherical harmonic and $%
\left\langle \Delta \psi \right\rangle =0$ we see that $\lambda =0$. Indeed,
for any $\psi $ we have $\lambda =0$ implies $\left\langle \tanh {\left[
\beta \left( \Omega L(x)+\lambda +\psi \right) \right] }\right\rangle =0$
and, 
\[
\frac d{d\lambda }\left\langle \tanh {\left[ \beta \left( \Omega
L(x)+\lambda +\psi \right) \right] }\right\rangle \\=\left\langle \frac
d{d\lambda }\tanh {\left[ \beta \left( \Omega L(x)+\lambda +\psi \right)
\right] }\right\rangle 
\]
\[
=\left\langle \beta \cosh ^{-2}{\left[ \beta \left( \Omega L(x)+\lambda
+\psi \right) \right] }\right\rangle 
\]
serves to show the average is monotonic in $\lambda $.

While the original discrete model was limited in its degrees of freedom, we
now have a continuum of freedom for the spin since now the coarse-grained
state is given by a bounded function $\sigma (x)$ in $L_2(S^2)$ with zero
circulation $\int_{S^2}dx\sigma (x)=0$. Further the Bragg estimation of
interaction energy became very accurate in the continuum model. The price
paid, like in all mean field models, is the artifact of the entropy
estimation. However, the Shannon entropy of a binary variable is exactly
what enables our derivation of the tanh fixed point equation above. The
figure below shows the spin states of some thermodynamic regimes obtained
numerically.

\begin{figure}[th]
\begin{center}
\includegraphics{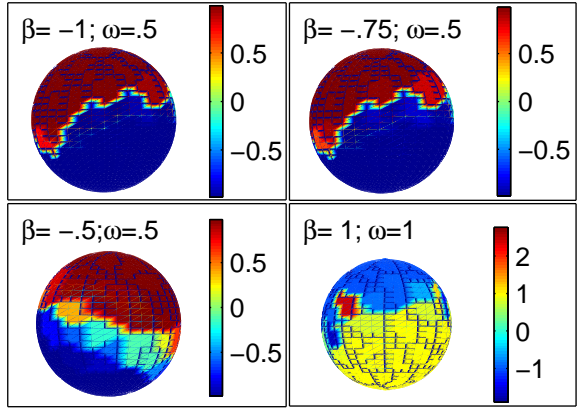}
\end{center}
\caption{Some fixed point solutions}
\end{figure}

\section{Conclusion}

In conclusion it must be emphasized that the combined angular momentum of
the fluid and solid sphere is conserved even when that of the fluid alone is
not. Similarly, without loss of generality for the purposes here, we can
assume that the complex torque mechanism responsible for coupling fluid to
sphere is conservative, and thus, the combined energy of the fluid-sphere
system is conserved.

In taking the mass of the rotating sphere to be infinite here and in \cite
{Limsphere06}, \cite{iutam06}, we have essentially modeled the statistical
mechanics of the coupled fluid-sphere problem as one having separate
reservoirs for angular momentum and energy. It was shown in \cite
{Limsphere06} and \cite{mean} that the particular form of the kinetic energy
functional for coupled barotropic flow on a sphere allows us to conveniently
combine these two reservoirs into one so there is only one inverse
temperature $\beta .$ Introducing a distinct reservoir for angular momentum,
that is, imposing a separate canonical constraint on the angular momentum of
the fluid will not change the physics but will add to the clutter in the
resulting model.

We remark here that a series of experiments on decaying and forced-damped
(or stationary) quasi-2D turbulence in square domains with no-slip and
stressless boundaries - see \cite{heijst2} and references therein - have
highlighted the role of variable angular momentum in inverse cascades to
self-organized structures. In particular, normal forces at the boundary of
the square can in principle produce a torque without requiring shear or the
presence of friction at the boundary. This conservative mechanism for
angular momentum transfer between fluid and boundary - despite the obvious
mismatch in geometry - provides the best experimental comparison that we
know to the (unresolved) torque mechanism introduced and developed by the
first author into the coupled fluid-sphere statistical mechanics models in a
series of recent papers.

For a more detailed view of the complex torque mechanism that couples the
fluid to the rotating sphere, one will have to invoke the observed and
theorized fact that a baroclinic component of the fluid forms Hadley cells
driven by non-uniform solar insolation which in turn serves to transfer high
momentum fluid in the equatorial region up to the middle atmosphere and
mid-latitudes. The planetary boundary layer is a very complex region that is
responsible for transferring the solid sphere's angular momentum to the
fluid as it flows towards the equator from mid-latitudes and the fluid's
angular momentum to the sphere between the polar region and mid-latitudes.
Any attempt to model this torque mechanism explicitly will result in
technical difficulties that can only be resolved by a huge computational
effort. Fortunately, the very essence of statistical mechanics - namely
coarse-graining - allows us to formulate a viable theory without explicitly
resolving this complex torque mechanism.

The results derived by using Bragg's method are in good agreement with those
found through Monte - Carlo simulations of the spherical energy-enstrophy
model by Ding and Lim \cite{Dinglim}, \cite{Dinglim2} and with a simple mean
field theory \cite{mean}, in the non-rotating, rotating positive temperature
and possibly rotating negative temperature problems. More details can be
found in the recent book \cite{limnebusbook} and on the website
http://www.rpi.edu/\symbol{126}limc. Further work should be done, however,
to refine the multi-dimensional Bragg method and extend it to more complex 
geophysical flows such as shallow water flows coupled to a rotating sphere.
The infinite-dimensional fixed point formulation in the previous section gave us
an interesting nonlinear elliptic equation (\ref{nleq2}) that should be further analyzed. 
Complementing the work here, the first author recently presented exact
solutions to the spherical model for the energy-relative enstrophy theory of
the coupled barotropic fluid-sphere system \cite{Limsphere06}, \cite{iutam06}%
.

\center{Acknowledgement}\\This work is supported by ARO grant
W911NF-05-1-0001 and DOE grant DE-FG02-04ER25616.

%\bibliography{BRAGGbib}

\end{document}